\documentclass[%
 reprint,
superscriptaddress,
 amsmath,amssymb,
 aps,
pra,
]{revtex4-1}

\usepackage{amsthm,amssymb}
\usepackage{amsmath}
\usepackage{graphicx}% Include figure files
\usepackage{dcolumn}% Align table columns on decimal point
\usepackage{bm}% bold math
\usepackage{multirow} % multirow
\usepackage{mathdots}
\usepackage{enumerate}
\usepackage{algorithm}
\usepackage{algorithmic}
\usepackage{color}
\usepackage{epsfig}
\usepackage{subfigure}
\usepackage{caption}
\theoremstyle{remark}

\begin{document}

\makeatletter
\newcommand{\ud}{\mathrm{d}}
\newcommand{\rmnum}[1]{\romannumeral #1}
\newcommand{\Rmnum}[1]{\expandafter\@slowromancap\romannumeral #1@}
\newcommand{\udots}{\mathinner{\mskip1mu\raise1pt\vbox{\kern7pt\hbox{.}}
        \mskip2mu\raise4pt\hbox{.}\mskip2mu\raise7pt\hbox{.}\mskip1mu}}
\makeatother

\preprint{APS/123-QED}

\title{Quantum algorithms for anomaly detection using amplitude estimation}

\author{Ming-Chao Guo}
\affiliation{State Key Laboratory of Networking and Switching Technology, Beijing University of Posts and Telecommunications, Beijing, 100876, China}
\affiliation{State Key Laboratory of Cryptology, P.O. Box 5159, Beijing, 100878, China}
\author{Hai-Ling Liu}
\affiliation{State Key Laboratory of Networking and Switching Technology, Beijing University of Posts and Telecommunications, Beijing, 100876, China}
\author{Yong-Mei Li}
\affiliation{State Key Laboratory of Networking and Switching Technology, Beijing University of Posts and Telecommunications, Beijing, 100876, China}
\author{Wen-Min Li}
\email{liwenmin@bupt.edu.cn}
\author{Su-Juan Qin}
\affiliation{State Key Laboratory of Networking and Switching Technology, Beijing University of Posts and Telecommunications, Beijing, 100876, China}
\author{Qiao-Yan Wen}
\affiliation{State Key Laboratory of Networking and Switching Technology, Beijing University of Posts and Telecommunications, Beijing, 100876, China}
\author{Fei Gao}
\email{gaof@bupt.edu.cn}
\affiliation{State Key Laboratory of Networking and Switching Technology, Beijing University of Posts and Telecommunications, Beijing, 100876, China}

\date{\today}

\begin{abstract}
Anomaly detection plays a critical role in fraud detection, health care, intrusion detection, military surveillance, etc. Anomaly detection algorithm based on density estimation (called ADDE algorithm) is one of widely used algorithms. Liang et al. proposed a quantum version of the ADDE algorithm [Phys. Rev. A 99, 052310 (2019)] and it is believed that the algorithm has exponential speedups on both the number and the dimension of training data point over the classical algorithm. In this paper, we find that Liang et al.'s algorithm doesn't actually execute. Then we propose a new quantum ADDE algorithm based on amplitude estimation. It is shown that our algorithm can achieves exponential speedup on the number $M$ of training data points compared with the classical counterpart. Besides, the idea of our algorithm can be applied to optimize the anomaly detection algorithm based on kernel principal component analysis (called ADKPCA algorithm). Different from the quantum ADKPCA proposed by Liu et al. [Phys. Rev. A 97, 042315 (2018)], compared with the classical counterpart, which offer exponential speedup on the dimension $d$ of data points, our algorithm achieves exponential speedup on $M$.
\end{abstract}

\pacs{Valid PACS appear here}
\maketitle

\section{Introduction}
Anomaly Detection (AD) refers to the problem of finding patterns in data that do not conform to expected behavior \cite{VC2009}. It is regarded as an essential branch in data mining and machine learning, and extensively used in a wide variety of fields, such as fraud detection for credit cards \cite{EB1997}, fault detection of safety-critical systems \cite{RT2005}, intrusion detection for network security \cite{VP2005} and health care \cite{SL2001}. Two common classical algorithms are AD algorithm based on Density Estimation \cite{MS2003} (ADDE algorithm) and AD algorithm based on Kernel Principal Component Analysis (ADKPCA algorithm) \cite{H2007}. In the presence of a large amount of input data, these two classical algorithms can be very computationally intensive. It is worthwhile to explore more effective algorithms for AD. 

Quantum computing has been demonstrated prominent advantages in some problems, such as factoring integers \cite{P1994}, searching in unstructured database \cite{L1996}, solving linear systems of equations \cite{AAS2009,LCS2018}, differential equation \cite{HYL2021}, quantum private query \cite{CXT2020,FSW2019,VSL2008}. The combination of quantum computing and machine learning created an emerging interdisciplinary field, Quantum Machine Learning (QML) \cite{JPN2017}, which is regarded as one of the most promising research directions. QML made great strides in data classification \cite{SMP2013,PMS2014,NDS2012}, clustering \cite{SBS2017}, neural network \cite{PTC2018}, linear regression \cite{G2017,CFQ2019,CFC2019}, association rule mining \cite{CFQ2016}, dimensionality reduction \cite{IC2016,SMP2014,SLH2020,CFS2019}, etc. 

In the context of quantum computing, several works have been developed to solve AD problems. In 2018, Liu et al. proposed a quantum ADKPCA algorithm \cite{NP2018}. The quantum algorithm calculates the inner product of two vectors based on swap-test \cite{HRJ2001,MIF2016} to obtain the value of proximity measure with complexity $O[$poly$(M\log d)]$, where $M$ and $d$ are the number and dimension of training data points, respectively. Subsequently, Liang et al. proposed a quantum version of the ADDE algorithm \cite{JSM2019} and claimed that the algorithm has exponential speedups on $M$ and $d$ compared with the classical counterpart.

In this paper, we find the mistakes of Liang et al.'s algorithm, which mainly derived from the failure of controlled rotation operation in the algorithm to extract classical information. This makes the whole algorithm doesn't actually execute. We propose a new quantum ADDE algorithm mainly utilizing amplitude estimation \cite{GPM2002}. Our algorithm can achieve exponential speedup over the classical algorithm on $M$. In addition, the idea of our algorithm can be applied to optimize the ADKPCA algorithm. Different from Liu et al.'s algorithm \cite{NP2018}, our algorithm shows exponential speedup on $M$ compared to the classical counterpart.

This paper is organized as follows. In Sec. \Rmnum{2}, we briefly review the classical ADDE and ADKPCA algorithms. In Sec. \Rmnum{3}, we present a new quantum ADDE algorithm and analyze its complexity in detail. In Sec. \Rmnum{4}, we give an application of the idea of our algorithm in the ADKPCA algorithm and analyze its complexity. The conclusion is given in Sec. \Rmnum{5}. The algorithm of Liang et al. is reviewed and analyzed in Appendix A and Appendix B, respectively.
\section{Review of ADDE and ADKPCA algorithms}
\label{sec:2}
In this section, we introduce the classical ADDE algorithm \cite{MS2003} and ADKPCA algorithm \cite{H2007}.

Given a training data set $\{\bm{x^i}\}_{i=1}^M$ consisting of $M$ normal data points and a new data point $\bm{x^0}$, where $\bm{x^i}=(x_1^i,\cdots,x_d^i)\in\mathcal{R}^d, i=0,1,\dots,M$. The traning data set can be represented by a $M\times d$ data matrix $X=(\bm{x}^1,\cdots,\bm{x}^M)^T$.
 \subsection{ADDE algorithm}
Assume that each feature is independent of each other and follows a Gaussian distribution $x_j^i\sim\mathcal{N}(\mu_j,\sigma_j^2),j=1,\cdots,d,i=1,\cdots,M$. The algorithm first establishes a statistical model $P(\bm{x})$ from the training data set. Then calculate the value of $P(\bm{x^0})$ and compare it with a pre-determined threshold $\delta$ to identify whether the new data point $\bm{x^0}$ is an anomaly or not. The whole procedure is depicted as follows.

(1) Fit the parameters mean $\mu_j$ and variance $\sigma_j^2$ by the training data set ($j=1,\cdots,d$):
\begin{equation}
		\mu_j=\frac{1}{M}\sum_{i=1}^Mx_j^i,\sigma_j^2=\frac{1}{M}\sum_{i=1}^M(x_j^i-\mu_j)^2.
\end{equation}

(2) Establish the statistical model $P(\bm{x})$ (the joint density function):
\begin{equation}
P(\bm{x})=\prod_{j=1}^dP(x_j;\mu_j,\sigma_j^2)=\prod_{j=1}^d\frac{1}{\sqrt{2\pi}\sigma_j}e^{-\frac{(x_j-\mu_j)^2}{2\sigma_j^2}}.
\end{equation}

(3) Given a new data point $\bm{x^0}$, calculate $P(\bm{x^0})$:
\begin{equation}
	P(\bm{x^{0}})=\prod_{j=1}^dP(x_j^0;\mu_j,\sigma_j^2)=\prod_{j=1}^d\frac{1}{\sqrt{2\pi}\sigma_j}e^{-\frac{(x_j^0-\mu_j)^2}{2\sigma_j^2}}\label{pythagorean}.	
\end{equation}
Simplify the Eq.(3):
\begin{align}
	\ln P(\bm{x^0})=-\frac{d}{2}\ln2\pi-\sum_{j=1}^d\ln\sigma_j-\sum_{j=1}^d\frac{(x_j^0-\mu_j)^2}{2\sigma_j^2}.	
\end{align}

Set a threshold $\delta$ in advance, if $\ln P(\bm{x}^0)\textless\ln\delta$, one can flag the new data as anomaly, otherwise, it is judged as normal. The complexity of ADDE algorithm is $O(Md)$. Therefore, when the size of the training data set is large, the ADDE algorithm may take a significant amount of time to execute.
\subsection{ADKPCA algorithm}

First, the mean vector $\bm{\mu}$ and covariance matrix $\Sigma$ of the training data set are defined as follows:
\begin{equation}
\bm{\mu}=\frac{1}{M}\sum_{i=1}^M\bm{x^i},\Sigma=\frac{1}{M-1}\sum_{i=1}^M\bm{(x^i-\mu)}(\bm{x^i-\mu})^T.
\end{equation}
The main purpose of the algorithm is to detect the difference in the distance between the new data point $\bm{x^0}$ and mean vector and the variance of the training data set along the direction $\bm{x^0-\mu}$. That is, to calculate the proximity measure:
\begin{equation}
	f(\bm{x^{0}})=|\bm{x^{0}-\mu}|^{2}-(\bm{x^{0}-\mu})^{T}\Sigma(\bm{x^{0}-\mu}).
\end{equation}
It can quantify how anomalous the point $\bm{x^0}$ is compared to the training data. A larger the value of $f(\bm{x^0})$ implies a more anomalous data point than a smaller $f(\bm{x^0})$. 

This method also allows us to classify anomalies in nonlinear feature spaces. The inner products are performed
in an abstract linear feature space. The inner product can be represented by a kernel function $k(\bm{x_i},\bm{x_j})$, which can be regarded as a non-linear function. For brevity, we only consider the linear kernel.

\section{A new quantum ADDE algorithm }
Liang et al.'s algorithm \cite{JSM2019} is reviewed in Appendix A. And we find the mistakes of this algorithm, which mainly derived from the failure of controlled rotation operation in the algorithm to extract classical information. This makes the whole algorithm doesn't actually execute. See Appendix B for a detailed analysis.

In this section, we present a new quantum ADDE algorithm in Sec. \Rmnum{3} A, and analyze its complexity in Sec. \Rmnum{3} B.

\subsection{Algorithm}

We assume that there are already quantum oracles that can efficiently access the elements in the data matrix $X$ and new data point $\bm{x^0}$ with time $O(\log Md)$ and $O(\log d)$, respectively. (for all $i=1,\cdots,M, j=1,\cdots,d.$) 
\begin{equation}
	\begin{aligned}
		&\bm{O_X}:|i\rangle|j\rangle|0\rangle\rightarrow|i\rangle|j\rangle|x_j^i\rangle,\\
		&\bm{O_x}:|j\rangle|0\rangle\rightarrow|j\rangle|x^0_j\rangle
	\end{aligned}
\end{equation}
This assumption can be made naturally when each element in $X$ and $\bm{x^0}$ is stored in quantum random access memory (QRAM)\cite{GSL2008}.

Our algorithm consists of three steps: in step 1, we prepare the state $\frac{1}{\sqrt{d}}\sum_{j=1}^d|j\rangle|\mu_j\rangle$; in step 2, the quantum state $\frac{1}{\sqrt{d}}\sum_{j=1}^d|j\rangle|\mu_j\rangle|\sigma_j^2\rangle|x_j^0\rangle$ can be generated; in step 3, we perform amplitude estimation \cite{GPM2002} to obtain the classic information of $\ln P(\bm{x^0})$. The entire algorithm process is shown in Fig. 1. 

In order to implement steps 1 and 2, we found that the parameter values of Eq. (1) can be regarded as the inner product of two vectors, ($j=1,2,\cdots,d$)
	\begin{align}
			&\mu_j=\frac{1}{M}(x_j^1,\cdots,x_j^M)\cdot(1,\cdots,1)^T,\nonumber\\
			&\sigma_j^2=\frac{1}{M}(x_j^1-\mu_j,\cdots,x_j^M-\mu_j)\cdot(x_j^1-\mu_j,\cdots,x_j^M-\mu_j)^T.
	\end{align}
These can be calculated in parallel by amplitude estimation and controlled rotation. 
\begin{figure*}[htb]
	\centering
		\includegraphics[height=10cm,width=11cm]{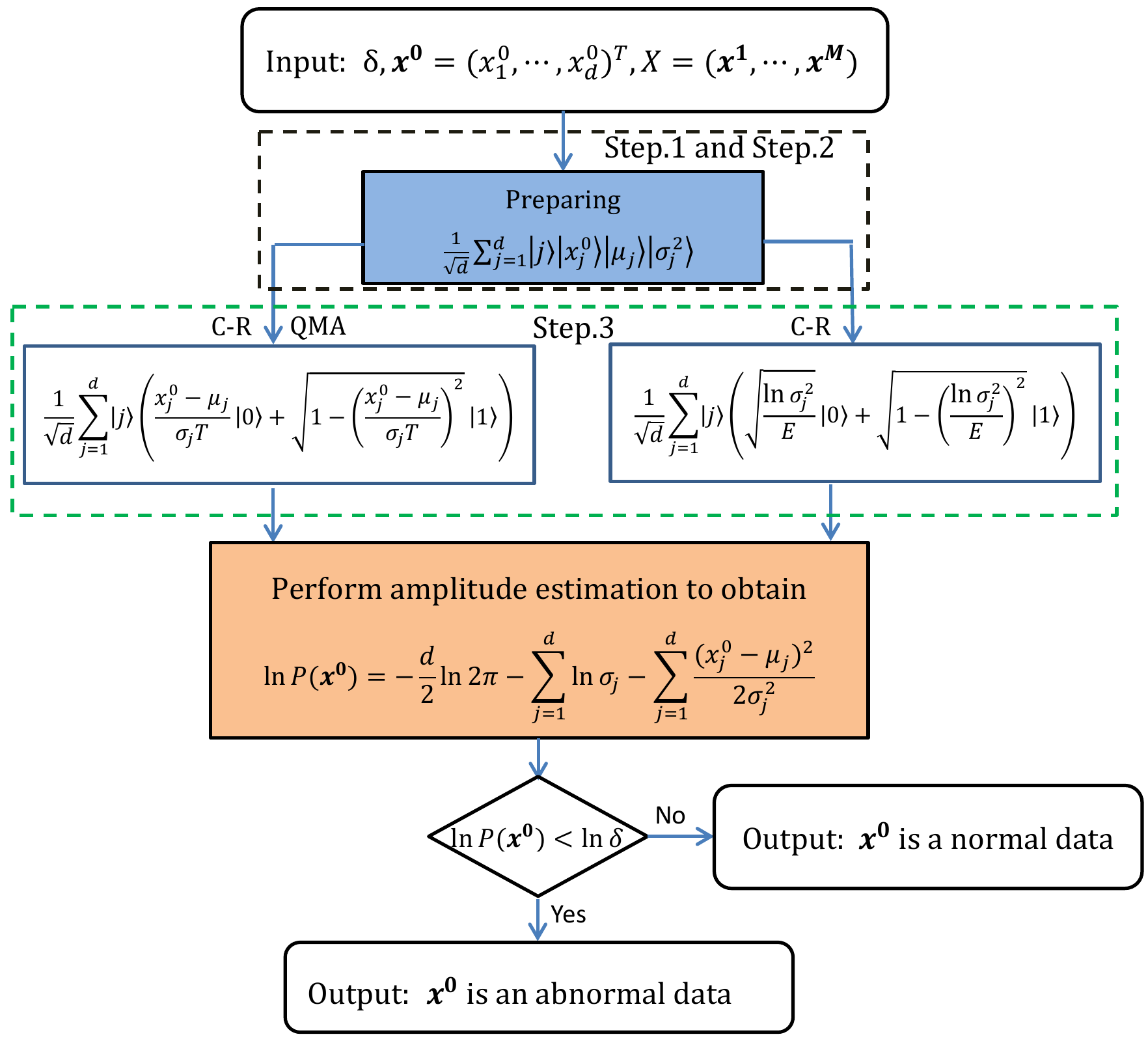}
\caption{The process diagram of quantum ADDE algorithm, where $\delta$ is a predetermined threshold, $C-R$ denotes controlled rotation and QMA denotes quantum multiply-adder.}
	\label{fig:kappa}
\end{figure*}

\textbf{\textit{Step 1.}} \textit{Prepare the state:} $ \frac{1}{\sqrt{d}}\sum_{j=1}^d|j\rangle|\mu_j\rangle$.

(1.1) Initialize quantum state 
\begin{equation}	
|0^{\otimes\log d}\rangle_1|0\rangle_2|0^{\otimes\log M}\rangle_3|0\rangle_4|0\rangle_5.
\end{equation}

(1.2) Perform $H^{\otimes\log d}, H$ and $H^{\otimes\log M}$ on the first, second and third register, respectively 
\begin{equation}
	\frac{1}{\sqrt{d}}\sum_{j=1}^d|j\rangle_1\frac{1}{\sqrt{2}}(|0\rangle+|1\rangle)_2\frac{1}{\sqrt{M}}\sum_{i=1}^M|i\rangle_3|0\rangle_4|0\rangle_5.
\end{equation} 

(1.3) Run the oracle $\bm{O_X}$ on the step (1.2) to prepare 
\begin{equation}
	\frac{1}{\sqrt{d}}\sum_{j=1}^d|j\rangle_1\frac{1}{\sqrt{2}}(|0\rangle+|1\rangle)_2\frac{1}{\sqrt{M}}\sum_{i=1}^M|i\rangle_3|x_j^i\rangle_4|0\rangle_5.
\end{equation}

(1.4) Perform $I_1\otimes(|0\rangle_2\langle0|_2\otimes I_3\otimes U_{45}+|1\rangle_2\langle1|_2\otimes I_3\otimes I_{45})$, where $U$ is a controlled rotation \cite{AAS2009, BJY2018}. Discard the fourth register, the state will becomes
\begin{equation}
	\begin{aligned}
	&\frac{1}{\sqrt{d}}\sum_{j=1}^d|j\rangle_1\frac{1}{\sqrt{2}}\bigg[|0\rangle\frac{1}{\sqrt{M}}\sum_{i=1}^M|i\rangle\bigg(\frac{x_j^i}{C}|0\rangle+\sqrt{1-(\frac{x_j^2}{C})^2}|1\rangle\bigg)+\\
	&|1\rangle\frac{1}{\sqrt{M}}\sum_{i=1}^M|i\rangle|0\rangle\bigg]_{235}:=\frac{1}{\sqrt{d}}\sum_{j=1}^d|j\rangle_1\frac{1}{\sqrt{2}}(|0\rangle|\phi_j\rangle+|1\rangle|h\rangle)_{235},
	\end{aligned}
\end{equation}
where $C=\max_{i,j}|x_j^i|$, $|h\rangle=\frac{1}{\sqrt{M}}\sum_{i=1}^M|i\rangle|0\rangle$ and $|\phi_j\rangle=\frac{1}{\sqrt{M}}\sum_{i=1}^M|i\rangle\bigg(\frac{x_j^i}{C}|0\rangle+\sqrt{1-(\frac{x_j^2}{C})^2}|1\rangle\bigg)$. 

(1.5) Apply $H$ gate to the second register
	\begin{align}
	&\frac{1}{\sqrt{d}}\sum_{j=1}^d|j\rangle_1\frac{1}{2}\big[|0\rangle_2(|\phi_j\rangle+|h\rangle)_{35}+|1\rangle_2(|\phi_j\rangle-|h\rangle)_{35}\big]\nonumber\\
	&:=\frac{1}{\sqrt{d}}\sum_{j=1}^d|j\rangle_1|\Phi_j\rangle_{235}.
	\end{align}
Each $|\Phi_j\rangle$ is rewritten as $|\Phi_j\rangle=\sin\theta_j|\Phi_j^0\rangle+\cos\theta_j|\Phi_j^1\rangle$, where $|\Phi_j^0\rangle, |\Phi_j^1\rangle$ are the normalized quantum states of $|0\rangle(|\phi_j\rangle+|h\rangle)$ and $|1\rangle(|\phi_j\rangle-|h\rangle)$, respectively. In addition,  $\sin^2\theta_j=\frac{1}{2}+\frac{1}{2}\langle\phi_j|h\rangle, \theta_j\in[0,\frac{\pi}{2}]$. We can use amplitude estimation \cite{34} to estimate $\langle\phi_j|h\rangle$. Define $Q_j=-A_jS_0A_j^{\dagger}S_{\chi}$, where $A_j:|0\rangle_{234}\rightarrow|\Phi_j\rangle, S_0=I-2|0\rangle_{234}\langle0|_{234}, S_{\chi}=(I-2|0\rangle_2\langle0|_2)\otimes I$. Perform $(Q_j)^l$ on the state $|\Phi_j\rangle$:
\begin{equation}
(Q_j)^l|\Phi_j\rangle=\sin[(2l+1)\theta_j]|\Phi_j^0\rangle+\cos[(2l+1)\theta_j]|\Phi_j^1\rangle,
\end{equation}
for any $l\in{\mathcal{N}}, Q_j$ acts as a rotation in the two subspace $\{|\Phi_j^0\rangle,|\Phi_j^1\rangle\}$, and it has two eigenvalues $e^{\pm\iota2\theta{j}}$ with eigenstates $|\Phi_j^{\uparrow,\downarrow}\rangle$ (un-normalized). 

(1.6) Add a ancillar register and excute amplitude estimation to generate
\begin{equation}
	\frac{1}{\sqrt{d}}\sum_{j=1}^d|j\rangle_1\bigg(|\Phi_j^{\uparrow}\rangle|\frac{\theta_j}{\pi}\rangle+|\Phi_j^{\downarrow}\rangle|\frac{1-\theta_j}{\pi}\rangle\bigg)_{2356}.
\end{equation}

(1.7) Compute  $|\mu_j\rangle=|C\cdot(2\sin^2{\theta_j}-1)\rangle$ via the quantum multiply-adder (QMA) and sine gate \cite{STJ2017,LJ2017}, 
\begin{equation}
	\frac{1}{\sqrt{d}}\sum_{j=1}^d|j\rangle_1|\widehat{\mu}_j\rangle_6,
\end{equation}
where $\widehat{\mu}_j$ represents the estimate of $\mu_j$. To simplify the notation, we add the tilde for estimates throughout this paper; that is, we denote $\widehat{\alpha}$ as the estimate of $\alpha$. The complete circuit diagram of step 1 is described in Figure 2.
\begin{figure}
	\centering
	\includegraphics[width=1\linewidth]{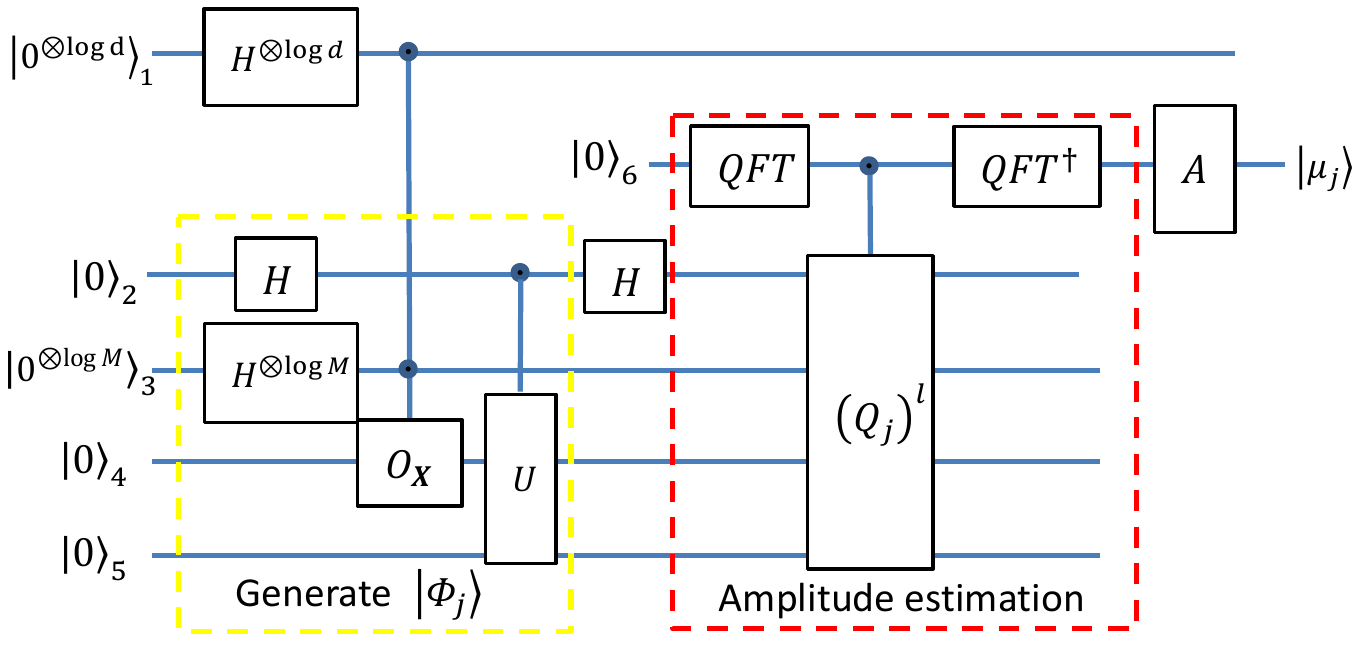}
	\caption{Quantum circuit diagram of step 1, where $A$ denotes the $C\cdot(2\sin^2(\pi\cdot)-1)$ gate to estimate $C\cdot\langle\bm{\phi}_j|\bm{h}\rangle$ and $QFT$ denotes quantum fourier transform.}
	\label{fig:2}
\end{figure}

\textbf{\textit{Step 2.}} \textit{Prepare the state:} $\frac{1}{\sqrt{d}}\sum_{j=1}^d|j\rangle|\mu_j\rangle|\sigma_j^2\rangle|x_j^0\rangle$.

(2.1) Prepare the quantum state 
\begin{equation}
 \frac{1}{\sqrt{d}}\sum_{j=1}^d|j\rangle_1|\widehat{\mu}_j\rangle_6|0\rangle_7|0\rangle_8.
 \end{equation}

(2.2) Perform $H^{\otimes\log M}$ gates on the seventh register and the oracle $\bm{O_X}$ to obtain
\begin{equation}
	\frac{1}{\sqrt{d}}\sum_{j=1}^d|j\rangle_1|\widehat{\mu}_j\rangle_6\frac{1}{\sqrt{M}}\sum_{i=1}^M|i\rangle_7|x_j^i\rangle_8.	
\end{equation} 

(2.3) The QMA gate is performed between sixth and eighth registers to generate 
\begin{equation}
	\frac{1}{\sqrt{d}}\sum_{j=1}^d|j\rangle_1|\widehat{\mu}_j\rangle_6\frac{1}{\sqrt{M}}\sum_{i=1}^M|i\rangle_7|x_j^i-\widehat{\mu}_j\rangle_8.
\end{equation}

(2.4) Add one qubit and run controlled rotation, discard the eighth register to generate
	\begin{align}
	&\frac{1}{\sqrt{d}}\sum_{j=1}^d|j\rangle_1|\widehat{\mu}_j\rangle_6\frac{1}{\sqrt{M}}\sum_{i=1}^M|i\rangle_7\bigg(\frac{x_j^i-\widehat{\mu}_j}{D}|0\rangle+\nonumber\\
	&\sqrt{1-\big(\frac{x_j^i-\widehat{\mu}_j}{D}\big)^2}|1\rangle\bigg)_9:=\frac{1}{\sqrt{d}}\sum_{j=1}^d|j\rangle_1|\widehat{\mu}_j\rangle_6|\chi_j\rangle_{79},
	\end{align}
where $D=\max_{i,j}|x_j^i-\mu_j|$.

(2.5) Similar to steps (1.5)-(1.6), we can get
\begin{equation}
\frac{1}{\sqrt{d}}\sum_{j=1}^d|j\rangle_1|\widehat{\mu}_j\rangle_6|\widehat{\sigma}_j^2\rangle_{10}.
\end{equation}

(2.6) Append an ancilla register and call the oracle $\bm{O_x}$
\begin{equation}
	\frac{1}{\sqrt{d}}\sum_{j=1}^d|j\rangle_1|\widehat{\mu}_j\rangle_6|\widehat{\sigma}_j^2\rangle_{10}|x_j^0\rangle_{f}.
\end{equation}
\textbf{\textit{Step 3.}} \textit{Compute} $\ln P(\bm{x^0})$. 

(3.1) Add one qubit to Eq. (22) and perform controlled rotation to obtain
\begin{equation}
	\begin{aligned}
		&\frac{1}{\sqrt{d}}\sum_{j=1}^d|j\rangle|\widehat{\mu}_j\rangle|\widehat{\sigma}_j^2\rangle|x_j^0\rangle\bigg(\frac{x_j^0-\widehat{\mu}_j}{\widehat{\sigma}_j T}|0\rangle+\sqrt{1-\big(\frac{x_j^0-\widehat{\mu}_j}{\widehat{\sigma}_j T}\big)^2}|1\rangle\bigg),	
	\end{aligned}
\end{equation}
where $T$ is a constant to ensure $\frac{x_j^0-\mu_j}{\sigma_j T}\le1$. 

(3.2) Run amplitude estimation to obtain
\begin{equation}
	\widehat{p}=\frac{1}{d}\sum_{j=1}^d\big(\frac{x_j^0-\widehat{\mu}_j}{\widehat{\sigma}_j T}\big)^2.
	\end{equation}

(3.3) According to step 2, we can prepare the state 
\begin{equation}
\frac{1}{\sqrt{2}}(|0\rangle+|1\rangle)_{e}\frac{1}{\sqrt{d}}\sum_{j=1}^d|j\rangle_1|\widehat{\sigma}_j^2\rangle_6|0\rangle_{g}
\end{equation} 

(3.4) Perform unitary operation $|0\rangle_e\langle0|_e\otimes I_1\otimes U_{6g}+|1\rangle_e\langle1|_e\otimes I_{16g}$. Uncompute the sixth register:
\begin{align}
 &\frac{1}{\sqrt{2}}\bigg[|0\rangle_e\frac{1}{\sqrt{d}}\sum_{j=1}^d|j\rangle_1\bigg(\frac{\ln\widehat{\sigma}_j^2}{E}|0\rangle+\sqrt{1-(\frac{\ln\widehat{\sigma}_j^2}{E})^2}|1\rangle\bigg)_g+\nonumber\\
 &|1\rangle_e\frac{1}{\sqrt{d}}\sum_{j=1}^d|j\rangle_1|0\rangle_g\bigg]:=\frac{1}{\sqrt{2}}(|0\rangle|\phi_j\rangle+|1\rangle|h'\rangle)_{e1g},
 \end{align}
where $|\phi_j\rangle:=\frac{1}{\sqrt{d}}\sum_{j=1}^d|j\rangle\bigg(\frac{\ln\widehat{\sigma}_j^2}{E}|0\rangle+\sqrt{1-(\frac{\ln\widehat{\sigma}_j^2}{E})^2}|1\rangle\bigg)$ and $|h'\rangle:=\frac{1}{\sqrt{d}}|1\rangle\sum_{j=1}^d|j\rangle|0\rangle$, $E=\max_j|\ln\sigma_j^2|$.

(3.5) Execute the $H$ gate to obtain
\begin{equation}
	\frac{1}{2}\big[|0\rangle_{e}(|\phi_j\rangle+|h'\rangle)_{1g}+|1\rangle_{e}(|\phi_j\rangle-|h'\rangle)_{1g}\big]
\end{equation}

(3.6) Run amplitude estimation to obtain 
\begin{equation}
		    \widehat{q}=\frac{1}{d}\sum_{j=1}^d\frac{\ln\widehat{\sigma}_j^2}{L}.	
\end{equation}

Hence, we can estimate the value of $\ln P(\bm{x^0})$, that is, 
\begin{equation}
\ln P(\bm{x^{0}})=-\frac{d}{2}\ln2\pi-\frac{1}{2}dL\cdot \widehat{q}-\frac{1}{2}dT^{2}\cdot \widehat{p}.	
\end{equation}

\subsection{Complexity Analysis}

(i). The complexity of step 1 is $O[(\log Md)/\varepsilon']$, which is mainly derived from the oracle $\bm{O_X}$ and amplitude estimation. Specific analysis is as follows: 

In steps (1.2)—(1.4), it takes $H$ gates, controlled rotation and oracle $\bm{O_X}$ with complexity $O(\log Md)$. In step (1.6), the amplitude estimation block needs $O(1/\varepsilon')$ applications of $Q_j=-A_jS_0A_j^{\dagger}S_{\chi}$ to achieve error $\varepsilon'$. The complexity of performing unitary operator $Q_j$ is $O(\log Md)$. In step (1.7), it takes QMA and sine gates with complexity $O($poly$\log1/\varepsilon')$, which is smaller than $O(1/\varepsilon')$, the complexity of these gates can be omitted. 

Now we analyze the error of step 1, which is mainly caused by the estimation of $\theta_j$, ($|\theta_j-\widehat{\theta}_j|\le\varepsilon'$)
\begin{equation}
	|\mu_j-\widehat{\mu}_j|=C|2\sin^2\theta_j-1-(2\sin^2\widehat{\theta}_j-1)|\le2C\varepsilon'.
\end{equation}
In a word, the complexity of \textit{step 1} is $O[(\log Md)/\varepsilon']$ and the error is $2C\varepsilon'$ to get $\frac{1}{\sqrt{d}}\sum_{j=1}^d|j\rangle|\widehat{\mu}_j\rangle$. 

(ii). The time complexity of step 2 is $O[(\log Md)/\varepsilon'\varepsilon'']$, which is mainly stem from the amplitude estimation of step (2.5). The detailed analysis is as follows: 

In step (2.1), perform the operations of \textit{step 1} with complexity $O[(\log Md)/\varepsilon']$. In step (2.2), it takes the oracle $\bm{O_X}$ with complexity $O(\log Md)$. In step (2.3), the QMA gate is performed and its complexity can be ignored.  In step (2.5), the amplitude estimation block needs $O(1/\varepsilon'')$ applications of the operation of step (2.4) to achieve error $\varepsilon''$. It implies the time complexity of step 2 is $O[(\log Md)/\varepsilon'\varepsilon'']$. In step (2.6), it takes the oracle $\bm{O_x}$ with complexity $O(\log d)$. The complexity of step 1 and step 2 are illustrated in the TABLE I.
\begin{table}[htbp]
	\captionsetup{singlelinecheck=off}
	\caption{The time complexity of step 1 and step 2}
	\begin{tabular}{ccccccc}
		\hline\hline
		steps&complexity&steps&complexity\\ \hline
		(1.2)-(1.4)&$O(\log Md)$&(2.1)&$O(\frac{\log Md}{\varepsilon'})$\\ 
		$-$& $-$ &(2.2)&$O(\log M+\log Md)$\\ 
		$-$& $-$ &(2.3)-(2.4)&$O($poly$\log \frac{1}{\varepsilon'})$\\ 
		(1.6)&$O(\frac{\log Md}{\varepsilon'})$&(2.5)&$O(\frac{\log Md}{\varepsilon'\varepsilon''})$\\
		(1.7)&$O($poly$\log\frac{1}{\varepsilon'})$&(2.6)&$O(\log d)$\\
		total&$O(\frac{\log Md}{\varepsilon'})$&total&$O(\frac{\log Md}{\varepsilon'\varepsilon''})$\\  \hline\hline
	\end{tabular}
\end{table}

Then we analyze the error of $\sigma_j^2$: 
\begin{align}
	&|\widehat{\sigma}_j^2-\sigma_j^2|=|\widehat{\sigma}_j^2-\frac{1}{M}\sum_{i=1}^M(x_j^i-\mu_j)^2|\nonumber\\
	&\le2D\cdot\varepsilon''+|\frac{1}{M}\sum_{i=1}^M(x_j^i-\widehat{\mu}_j)^2-\frac{1}{M}\sum_{i=1}^M(x_j^i-\mu_j)^2|\nonumber \\
	&\le2D\cdot\varepsilon''+8C^2\varepsilon',
\end{align}
where $2D\cdot\varepsilon''$ is derived from the amplitude estimation in step (2.5).

(iii). The complexity of step 3 is mainly came from amplitude estimation of step (3.2) and step (3.6), and assume its error is $\varepsilon'''$. The complexity of estimating the values of $\sum_{j=1}^d(\frac{x_j^0-\mu_j}{\sigma_j})^2$ and $\sum_{j=1}^d\ln\sigma_j$ are $O(\frac{\log Md}{\varepsilon'\varepsilon''\varepsilon'''})$, and their errors are as shown:
\begin{align}
		&|dE\widehat{q}-\sum_{j=1}^M\ln\sigma_j^2|=E|d\widehat{q}-\sum_{j=1}^d\frac{\ln\widehat{\sigma}_j^2}{E}+\frac{\ln\widehat{\sigma}_j^2}{E}-\frac{\ln\sigma_j^2}{E}|\nonumber\\
		&\le E(d\varepsilon'''+|\sum_{j=1}^d\frac{\ln\widehat{\sigma}_j^2-\ln\sigma_j^2}{E}|)\\
	    &\le E\varepsilon'''+|\ln(1+\frac{8C^2\varepsilon'+2D\varepsilon''}{\min\sigma_j})|\le\frac{(E+3\min\sigma_j)\varepsilon}{3T^2},\nonumber
	    \end{align}
\begin{align}
	   &|dT^2\widehat{p}-\sum_{j=1}^d\frac{(x_j^0-\mu_j)^2}{\sigma_j^2}|\nonumber\\
	   &=dT^2|\widehat{p}-\sum_{j=1}^d\frac{(x_j^0-\widehat{\mu}_j)^2}{dT^2\widehat{\sigma}_j^2}+\sum_{j=1}^d\frac{(x_j^0-\widehat{\mu}_j)^2}{dT^2\widehat{\sigma}_j^2}-\frac{(x_j^0-\mu_j)^2}{dT^2\sigma_j^2}|\nonumber\\
	   &\le dT^2\varepsilon'''+|\sum_{j=1}^d\frac{\sigma_j^2(x_j^0-\widehat{\mu}_j)^2-\widehat{\sigma}_j^2(x_j^0-\mu_j)^2}{\widehat{\sigma}_j^2\sigma_j^2}|\\
	   &\le dT^2\varepsilon'''+dT^2\frac{8C^2\varepsilon'+2D\varepsilon''}{\min_{j}\sigma_j^2}+d\frac{8C^2\varepsilon'}{\min_{j}\sigma_j^2}\nonumber.	
\end{align}
If $C,D,T,E,\min_{j}\sigma_j^2=O(1)$, and $\varepsilon'''=\frac{\varepsilon}{3dT^2}, \varepsilon''=\frac{\min_j\sigma_j^2\varepsilon}{3dT^2D}, \varepsilon'=\frac{\min_j\sigma_j^2\varepsilon}{3d(8T^2C^2+8C^2)}.$ The overall
runtime will be 
\begin{align}
	O[d^3\log (Md)\varepsilon^{-3}]\nonumber.
\end{align}
That is, we can get an $\varepsilon$-approximate of $\ln P(\bm{x^0})$ in complexity $O[d^3\log(Md)\varepsilon^{-3}]$.
This means that our quantum algorithm is exponentially faster than the corresponding classical algorithm
when $1/\varepsilon, d=O(\log M)$.

\section{application in ADKPCA algorithm}

In this section, we propose an application of the idea of our algorithm in the ADKPCA algorithma, and assume that there are already quantum oracles $\bm{O_X}$ and $\bm{O_x}$,  which are the same as the quantum ADDE algorithm. 
\subsection{Algorithm}
In order to calculate the proximity measure $f(\bm{x^0})$, our quantum algorithm consists of the following two steps:\\
\textbf{\textit{Step 1.}} \textit{Computing} $|\bm{x^{0}-\mu}|^{2}$.

(1.1) According to step 1 in Sce. III A, we can prepare quantum state $\frac{1}{\sqrt{d}}\sum_{j=1}^d|j\rangle|0\rangle|\widehat{\mu}_j\rangle$.

(1.2) Perform the oracle $\bm{O_x}$ on step (1.1): 
\begin{equation}
\frac{1}{\sqrt{d}}\sum_{j=1}^d|j\rangle|0\rangle|\widehat{\mu}_j\rangle\stackrel{\bm{O_x}}{\longrightarrow}\frac{1}{\sqrt{d}}\sum_{j=1}^d|j\rangle|x^0_j\rangle|\widehat{\mu}_j\rangle.
\end{equation}

(1.3) Excute QMA gate and controlled rotation to generate:
\begin{align}
	&\frac{1}{\sqrt{d}}\sum_{j=1}^d|j\rangle|x_j^0-\widehat{\mu}_j\rangle(\frac{x_j^0-\widehat{\mu}_j}{C'}|0\rangle+\sqrt{1-(\frac{x_j^0-\widehat{\mu}_j}{C'})^2}|1\rangle)\nonumber\\
	&:=|\bm{\psi_0\rangle}+|\bm{\psi_1\rangle},
\end{align}
where $C'=\max_j|x_j^0-\mu_j|$.

(1.4) Run amplitude estimation, we can get:
\begin{equation}
	\widehat{a}=\frac{1}{d}\sum_{j=1}^d(\frac{x_j^0-\widehat{\mu}_j}{C'})^2.
\end{equation}
Therefore, $|\bm{z^0}|^2=\sum_{j=1}^d(x_j^0-\mu_j)^2=d(C')^2\cdot\widehat{a}$.

\textbf{\textit{Step 2.}} \textit{Computing} $(\bm{x^{0}-\mu})^{T}\Sigma(\bm{x^{0}-\mu})$.

(2.1) Prepare the state $|0\rangle|0\rangle\frac{1}{\sqrt{d}}\sum_{j=1}^d|j\rangle|x_j^0\rangle|\widehat{\mu}_j\rangle$.

(2.2) Perform $H$ gates and the oracle $\bm{O_X}$ to obtain:
\begin{equation}
	\frac{1}{\sqrt{M}}\sum_{i=1}^M|i\rangle|x_j^i\rangle\frac{1}{\sqrt{d}}\sum_{j=1}^d|j\rangle|x_j^0\rangle|\widehat{\mu}_j\rangle
\end{equation}

(2.3) Perform QMA gate and control rotation operation, uncompute the last three registers, obtain
\begin{align}
	&\frac{1}{\sqrt{M}}\sum_{i=1}^M|i\rangle\frac{1}{\sqrt{d}}\sum_{j=1}^d|j\rangle\bigg(\frac{(x_j^i-\widehat{\mu}_j)(x_j^0-\widehat{\mu}_j)}{C''}|0\rangle\nonumber\\
	&+\sqrt{1-\big(\frac{(x_j^i-\widehat{\mu}_j)(x_j^0-\widehat{\mu}_j)}{C''}\big)^2}|1\rangle\bigg):=\frac{1}{\sqrt{M}}\sum_{i=1}^M|i\rangle|\Psi_i\rangle,
\end{align} 
where $C''=\max_{i,j}|(x_j^i-\mu_j)(x_j^0-\mu_j)|$.

(2.4) Similar to step (1.3) in Sec. III A, we can prepare 
\begin{align}
	&\frac{1}{\sqrt{M}}\sum_{i=1}^M|i\rangle\frac{1}{2}[|0\rangle(|\Psi_i\rangle+|h'\rangle)+|1\rangle(|\Psi_i\rangle-|h'\rangle)]\nonumber\\
	&:=\frac{1}{\sqrt{M}}\sum_{i=1}^M|i\rangle|\varphi_i\rangle,
\end{align}
where $|h'\rangle=\frac{1}{\sqrt{d}}\sum_{j=1}^d|j\rangle|0\rangle$.

(2.5) Excute amplitude estimation, QMA and sine gate (similar to the steps (1.6)-(1.7) in Sec. III A), generate 
\begin{equation}
	\frac{1}{\sqrt{M}}\sum_{i=1}^M|i\rangle|\widehat{\omega}_i\rangle,
\end{equation}
where $\omega_i:=\langle\Psi_i|h'\rangle=\frac{1}{d}\sum_{j=1}^d\frac{(x_j^0-\widehat{\mu}_j)(x_j^i-\widehat{\mu}_j)}{C''}$.

(2.6) Add an ancilla register and perform controlled rotation to obtain the quantum state:\begin{equation}
	\frac{1}{\sqrt{M}}\sum_{i=1}^M|i\rangle|\widehat{\omega}_i\rangle(\widehat{\omega}_i|0\rangle+\sqrt{1-\widehat{\omega}_i^2}|1\rangle).
\end{equation}

(2.7) Run amplitude estimation to obtain
\begin{equation}
	\widehat{b}=\frac{1}{M}\sum_{i=1}^M(\widehat{\omega}_i)^2.
\end{equation}
The value of Eq. (6) can be approximated as 
\begin{equation}
	f(\bm{x^{0}})=d(C')^2\cdot\widehat{a}-\frac{M}{M-1}(dC'')^2\cdot\widehat{b}.
\end{equation}
\subsection{Complexity analysis}
(i). The complexity of step 1 is $O(\frac{\log Md}{\varepsilon'\varepsilon''})$, which is mainly derived from the amplitude estimation. The detailed analysis is shown below:

According to step 1 performed in Sce. III A, the time complexity of step (1.1) is $O(\frac{\log Md}{\varepsilon'})$, and error is $2C\epsilon'$. In step (1.2), the oracle $\bm{O_x}$ is performed with complexity $O(\log d)$. In step (1.3), it takes QMA and controlled rotation with complexity $O[$poly$\log(1/\epsilon')]$, the complexity of these gates can be omitted. In step (1.4), the amplitude estimation is performed with complexity $O(\frac{\log Md}{\varepsilon'\varepsilon''})$ to achieve error $\varepsilon''$. 
\begin{equation}
	|\widehat{a}-\frac{1}{d}\sum_{j=1}^d(\frac{x_j^0-\widehat{\mu}_j}{C'})^2|\le\varepsilon''.
\end{equation}
Therefore, the error of step 1 is shown below
\begin{align}
	&|d(C')^2\cdot\widehat{a}-\sum_{j=1}^d(x_j^0-\mu_j)^2|\nonumber\\
	&\le d(C')^2\epsilon''+|\sum_{j=1}^d(x_j^0-\widehat{\mu}_j)^2-\sum_{j=1}^d(x_j^0-\mu_j)^2|\notag \\
	&\le d(C')^2\epsilon''+4dC'C\epsilon',
\end{align}
where $|\mu_j-\widehat{\mu}_j|\le2C\varepsilon'$ and $C=\max_{i,j}|x_j^i|$, which is the same as the error analysis of Step 1 in Sce. III A. When $\varepsilon'=\frac{\varepsilon}{48d^2C''}$ and $\varepsilon''=\frac{\varepsilon}{d(C'')^2}$, $|d(C')^2\cdot a-\sum_{j=1}^d(x_j^0-\mu_j)^2|\le\varepsilon$. Then, step 1 returns an $\varepsilon$-approximate of $|\bm{x^0-\mu}|^2$ in complexity $O[d^3\log(Md)\varepsilon^{-3}]$.

(ii). The complexity of step 2 is $O(\frac{\log Md}{\varepsilon'\epsilon'''\epsilon''''})$. The specific analysis is as follows:

In step (2.1), perform the Oracle $\bm{O_X}$ with complexity of $O(\log Md)$. In steps (2.2)-(2.3), the $H$ gate, QMA and controlled rotation are performed. In step (2.4), we still need the information of $\omega_i$ to do further operations, and thus we store it in an ancilla register as Eq. (40). But, in step (2.6), we need an output about the estimate of $\frac{1}{M}\sum_{i=1}^M\omega_i^2$. Hence, after the amplitude estimation, we need to perform a measurement on the auxiliary register. It is assumed that the errors of amplitude estimation of step (2.4) and step (2.6) are $\epsilon'''$ and $\epsilon''''$, respectively. The complexity of solving the value of $(\bm{z^{0}})^{T}\Sigma(\bm{z^{0}})$ is $O(\frac{\log Md}{\varepsilon'\epsilon'''\epsilon''''})$.The error is shown below:
\begin{widetext}
	\begin{eqnarray}
		\bigg|\widehat{b}-b+b-\frac{1}{Md^2}\sum_{i=1}^M{\big[\sum_{j=1}^d\frac{(x_j^0-\widehat{\mu}_j)(x_j^i-\widehat{\mu}_j)}{C''}\big]^2+\big[\sum_{j=1}^d\frac{(x_j^0-\widehat{\mu}_j)(x_j^i-\widehat{\mu}_j)}{C''}\big]^2-\big[\sum_{j=1}^d\frac{(x_j^0-\mu_j)(x_j^i-\mu_j)}{C''}\big]^2}\bigg|,
	\end{eqnarray}
\end{widetext}
where the difference of the first two items is the error of the amplitude estimation of step (2.6), and the difference of the third and fourth items is the error of the amplitude estimation of step (2.4). That is, 
\begin{equation}
	|\widehat{b}-\frac{1}{M}\sum_{i=1}^M\big[\sum_{j=1}^d\frac{(x_j^0-\mu_j)(x_j^i-\mu_j)}{dC''}\big]^2|\le\varepsilon''''+\varepsilon'''+\frac{16C^2\varepsilon'}{C''}.
\end{equation}
When $\varepsilon''''=\frac{\varepsilon}{3d^2(C'')^2},\varepsilon'''=\frac{\varepsilon}{3d^2(C'')^2}$ and $\varepsilon'=\frac{\varepsilon}{48d^2C''}$, $|(dC'')^2\cdot \widehat{b}-\frac{1}{M-1}\big[\sum_{j=1}^d(x_j^0-\mu_j)(x_j^i-\mu_j)\big]^2|\le\varepsilon$. That is, the algorithm returns an $2\varepsilon$-approximate of $f(\bm{x^0})$ in complexity $O[d^6\log(Md)\varepsilon^{-3}]$. 

Different from Liu et al's algorithm \cite{NP2018} calculates inner product of two vectors based on swap-test, our algorithm calculates $M$ inner product values in parallel based on the amplitude estimation. This means that our algorithm achieves exponential speedup on $M$ compared with the classical counterpart.

\section{Conclusion}
In the present study, we found the mistakes in the Liang et al.'s algorithm and proposed a new quantum ADDE algorithm mainly based on amplitude estimation. Our algorithm can achieve exponential speedup on the number of training data points $M$ compared with the classical counterpart. Moreover, our algorithm can be applied to optimize the ADKPCA algorithm. This quantum algorithm has exponential speed on $M$ compared with the classical algorithm, different from Liu et al.'s algorithm. We hope our algorithms and particularly the key technique used in our algorithms, amplitude estimation,  can inspire more efficient quantum machine learning algorithms.

\section*{Acknowledgements}
We thank Shijie Pan and Linchun Wan for useful discussions on the subject. This work is supported by the Fundamental Research Funds for the Central Universities (Grant No.2019XD-A01) and NSFC (Grants No.61976024, No.61972048).
\appendix
\section{Liang et al.'s algorithm}
Given a set $\{|\bm{x^i}\rangle\}_{i=1}^M$ composed of $M$ normal quantum training states, where $|\bm{x^i}\rangle=\sum_{j=1}^dx_j^i|j\rangle,$ and $\bm{x^i}=(x_1^i,\cdots,x_d^i)^T,i=1,\cdots,M $ is a normalized vector. The objective is to detect how anomalous the new quantum state $|\bm{x^0}\rangle$ is compared to the normal states. The algorithm is as follows:

Given the control unitaries to create the following superpositions of training states
\begin{equation}
	\frac{|0\rangle\sum_{i=1}^M|\bm{x^i}\rangle|i\rangle+|1\rangle|\bm{\mu}\rangle\sum_{i=1}^M|i\rangle}{\sqrt{2}},
\end{equation}
where $|\bm{\mu}\rangle=\frac{1}{N_{\mu}}\sum_{j=1}^d\mu_j|j\rangle,\mu_j=\sum_{i=1}^Mx_j^i$, $N_{\mu}$ is normalized coefficient.

Apply $H$ gate on qubit of the first register to obtain
\begin{equation}
	\dfrac{1}{2}\big[|0\rangle\sum_{j=1}^d\sum_{i=1}^M(x_j^i+\mu_j)|i\rangle|j\rangle+|1\rangle\sum_{j=1}^d\sum_{i=1}^M(x_j^i-\mu_j)|i\rangle|j\rangle\big].
\end{equation}
Measure the first register in $|1\rangle$, the remaining qubits collapse into the state
\begin{equation}
	\sum_{j=1}^d\sum_{i=1}^M(x_j^i-\mu_j)|i\rangle|j\rangle=\sum_{j=1}^d|\chi_j\rangle|j\rangle.
\end{equation}
Prepare quantum states $\sum_{j=1}^d\sum_{i=1}^M(x_j^0-\mu_j)|i\rangle|j\rangle=\sum_{j=1}^d|\chi_j^0\rangle|j\rangle$ by the same method.\\
\textbf{\textit{Algorithm 1.}} \textit{Efficiently computing} $\sum_{j=1}^d\frac{(x_j^0-\mu_j)^2}{2\sigma_j^2}$.

(1.1) Prepare the quantum state
\begin{equation}
	\sum_{j=1}^d|\chi_j\rangle|\chi_j^0\rangle|j\rangle,
\end{equation}
where $|\chi_j\rangle=\sum_{i=1}^M(x_j^i-\mu_j)|i\rangle$, $\mu_j=\frac{1}{M}\sum_{i=1}^Mx_j^i$ and $|\chi_j^0\rangle=\sum_{i=1}^M(x_j^0-\mu_j)|i\rangle$.

(1.2) Add an ancilla qubit $|0\rangle$ and perform a controlled unitary operator $R_1$ to obtain\\
\begin{equation}
	\sum_{j=1}^d|\chi_j\rangle|\chi_j^0\rangle|j\rangle(\frac{\chi_j^0}{\chi_j}|0\rangle+\sqrt{1-(\frac{\chi_j^0}{\chi_j})^2}|1\rangle)	
\end{equation}

(1.3) Uncompute the second and third registers, the system state is
\begin{equation}
	\sum_{j=1}^d|j\rangle(\frac{\chi_j^0}{\chi_j}|0\rangle+\sqrt{1-(\frac{\chi_j^0}{\chi_j})^2}|1\rangle)	
\end{equation}

(1.4) Measure the quantum observation  $M_1=I\otimes|0\rangle\langle0|$ and the expectation $\langle M_1\rangle=\sum_{j=1}^d(\frac{\chi_j^0}{\chi_j})^2$.\\
\textbf{\textit{Algorithm 2.}} \textit{Efficiently computing $\sum_{j=1}^d\ln\sigma_j$}

(2.1) Prepare the quantum state $\sum_{j=1}^d|\chi_j\rangle|j\rangle$.

(2.2) Add an ancilla qubit $|0\rangle$ and perform a controlled unitary operator $R_2$ to obtain
\begin{equation}
	\sum_{j=1}^d|\chi_j\rangle|j\rangle(\ln\chi_j|0\rangle+\sqrt{1-2\ln\chi_j}|1\rangle)
\end{equation}

(2.3) Uncompute the second register. The system state becomes
\begin{equation}
	\sum_{j=1}^d|j\rangle(\ln\chi_j|0\rangle+\sqrt{1-2\ln\chi_j}|1\rangle)
\end{equation}

(2.4) Measure the quantum observation  $M_2=I\otimes|0\rangle\langle0|$ and the expectation $\langle M_2\rangle=2\sum_{j=1}^d\ln\chi_j$.

If $\ln P(|\bm{x^0}\rangle)=-\frac{d}{2}\ln2\pi-\frac{1}{2}(\langle M_1\rangle+\langle M_2\rangle)<\ln\delta$, one can flag $|\bm{x^0}\rangle$ as an anomaly quantum state. Otherwise, it is normal.

\section{Analysis of Liang et al.'s algorithm}
To facilitate understanding, we first introduce controlled rotation \cite{AAS2009, BJY2018} and two forms of quantum states corresponding to classical data: digital-encoded state and analog-encoded state \cite{KMK2019}, as shown below:

(a) Analog-encoded state: Let $\bm{\alpha}=(\alpha_1,\cdots,\alpha_N)$ denote a normalized vector, the quantum state $\sum_{j=1}^N\alpha_j|j\rangle$ is called an analog-encoded state. 

(b) Digital-encoded state: Let $\beta=\beta_1\cdots\beta_n$ is an $n$-bit real number, the quantum state $|\beta_1\cdots\beta_n\rangle$ is called a digital-encoded state.

(c) Controlled rotation: Let $x$ denote an $n$-bit real number, there exist unitary operation $U_{c}$ satisfying 	
\begin{equation}
	|x\rangle|0\rangle\stackrel{U_{c}}{\longrightarrow}|x\rangle(f(x)|0\rangle+\sqrt{1-f(x)^2}|1\rangle),
\end{equation}
where $|x\rangle$ is a digital-encoded state, $|f(x)|\le1$ and the function $f$ can be computed efficiently on a classical computer.

Then, we analyze and point out the mistakes of Liang et al.'s algorithm, as shown below:

1. In this algorithm, it is claimed that the controlled rotation operations $R_1$ and $R_2$ can be performed:
\begin{align}
	&\sum_{j=1}^d|\chi_j\rangle|\chi_j^0\rangle|0\rangle\stackrel{R_1}{\longrightarrow}|\chi_j\rangle|\chi_j^0\rangle\big(\frac{\chi_j^0}{\chi_j}|0\rangle+\sqrt{1-(\frac{\chi_j^0}{\chi_j})^2}|1\rangle\big),\nonumber\\ &\sum_{j=1}^d|\chi_j\rangle|0\rangle\stackrel{R_2}{\longrightarrow}|\chi_j\rangle(\ln\chi_j|0\rangle+\sqrt{1-(\ln\chi_j)^2}|1\rangle),
\end{align}
where $|\chi_j\rangle=\sum_{i=1}^M(x_j^i-\mu_j)|i\rangle$, $|\chi_j^0\rangle=\sum_{i=1}^M(x_j^0-\mu_j)|i\rangle$. However, $|\chi_j\rangle$ and $|\chi_j^0\rangle$ are not digital-encoded state, and $x_j^i-\mu_j$ is unknown in advance ($i=0,1,\cdots,M,j=1,\cdots,d$). This leads to the $R_1$ and $R_2$ cannot be effectively implemented. 

2. It is claimed the quantum states $\sum_{j=1}^d\sum_{i=1}^M(x_j^i-\mu_j)|i\rangle|j\rangle$ and $\sum_{j=1}^d\sum_{i=1}^M(x_j^0-\mu_j)|i\rangle|j\rangle$ can be prepared (See Eq. (A1)—(A3) in Appendix A). However,  $|\bm{\mu}\rangle=\frac{1}{N_{\mu}}\sum_{j=1}^d\mu_j|j\rangle$, and performing the $H$ gate on Eq. (A1) to generate
\begin{equation}
	\frac{1}{2}\big[|0\rangle\sum_{j=1}^d\sum_{i=1}^M(x_j^i+\frac{\mu_j}{N_{\mu}})|i\rangle|j\rangle+|1\rangle\sum_{j=1}^d\sum_{i=1}^M(x_j^i-\frac{\mu_j}{N_{\mu}})|i\rangle|j\rangle\big].
\end{equation}
Then measure the first register in $|1\rangle$, the Eq. (9) collapse into the state $\sum_{j=1}^d\sum_{i=1}^M(x_j^i-\frac{\mu_j}{N_{\mu}})|i\rangle|j\rangle$, which is not consistent with the quantum state of Eq. (A3). Similarly, the state $\sum_{j=1}^d\sum_{i=1}^M(x_j^0-\mu_j)|i\rangle|j\rangle$ cannot be obtained.

3. It is believed that $\langle M_2\rangle=2\sum_{j=1}^d\ln\chi_j$. In fact, 
\begin{equation}
	\langle M_2\rangle=\sum_{j=1}^d(\ln\chi_j)^2\neq2\sum_{j=1}^d\ln\chi_j\neq\sum_{j=1}^d\ln\sigma_j^2.
\end{equation}
This causes the algorithm failing to reach its goal, i.e,
\begin{equation}
	\ln P(|\bm{x^0}\rangle)\neq-\frac{d}{2}\ln2\pi-\frac{1}{2}(\langle M_1\rangle+\langle M_2\rangle).
\end{equation}

\end{document}